\documentclass[a4paper,12pt]{article}

\usepackage{amsmath}
\usepackage[psamsfonts]{amssymb}
\usepackage{rsfs}
\usepackage{bm}

\usepackage{cite}
\usepackage[dvips]{graphicx}
\usepackage{color}

\begin{document} 

\begin{titlepage}

\baselineskip 10pt
\hrule 
\vskip 5pt
\leftline{}
\leftline{Chiba Univ./KEK Preprint
          \hfill   \small \hbox{\bf CHIBA-EP-165}}
\leftline{\hfill   \small \hbox{\bf KEK Preprint 2007-19}}
\leftline{\hfill   \small \hbox{July 2007}}
\vskip 5pt
\baselineskip 14pt
\hrule 
\vskip 0.5cm

\centerline{\Large\bf 
Compact lattice formulation of 
} 
\vskip 0.3cm
\centerline{\Large\bf  
Cho-Faddeev-Niemi decomposition: 
}
\vskip 0.3cm
\centerline{\Large\bf  
gluon mass generation and infrared Abelian dominance 
}
\centerline{\large\bf  
}

\vskip 0.3cm

\centerline{{\bf 
A. Shibata$^{\flat,{1}}$,  
S. Kato$^{\sharp,{2}}$, 
K.-I. Kondo$^{\dagger,\ddagger,{3}}$,  
T. Murakami$^{\ddagger,{4}}$,
T. Shinohara$^{\ddagger,{5}}$
$\&$
S. Ito$^{\star,{6}}$ 
}}  
\vskip 0.5cm
\centerline{\it
${}^{\flat}$%
Computing Research Center, High Energy Accelerator Research Organization (KEK),  
}
\vskip 0.1cm
\centerline{\it
\& 
Graduate Univ. for Advanced Studies (Sokendai),
Tsukuba 
305-0801, 
Japan
}
\vskip 0.3cm
\centerline{\it
${}^{\sharp}$Takamatsu National College of Technology, Takamatsu 761-8058, Japan
}
\vskip 0.3cm
\centerline{\it
${}^{\dagger}$Department of Physics, Faculty of Science, 
Chiba University, Chiba 263-8522, Japan
}
\vskip 0.3cm
\centerline{\it
${}^{\ddagger}$Graduate School of Science and Technology, 
Chiba University, Chiba 263-8522, Japan
}
\vskip 0.3cm
\centerline{\it
${}^{\star}$Nagano National College of Technology, 716 Tokuma, Nagano 381-8550, Japan
}
\vskip 0.3cm
\vskip 0.1cm
\centerline{\it
}
\vskip 0.3cm
\vskip 0.5cm

\begin{abstract}
This paper complements a new lattice formulation of SU(2) Yang-Mills theory written in terms of new variables in a compact form proposed in  the previous paper.  
The new variables used in the formulation were once called the Cho--Faddeev--Niemi or Cho--Faddeev--Niemi--Shabanov decomposition. 
Our formulation enables us to explain the infrared ``Abelian'' dominance, in addition to magnetic monopole dominance shown in the previous paper,  in the gauge invariant way without relying on the specific gauge fixing called the maximal Abelian gauge used in the conventional investigations.  In this paper, especially, we demonstrate by numerical simulations that gluon degrees of freedom other than the ``Abelian'' part acquire the mass to be decoupled in the low-energy region leading to the infrared Abelian dominance.
\end{abstract}


\hrule  
\vskip 0.1cm
${}^1$ 
  E-mail:  {\tt akihiro.shibata@kek.jp}
  
${}^2$ 
  E-mail:  {\tt kato@takamatsu-nct.ac.jp}
  
${}^3$ 
  E-mail:  {\tt kondok@faculty.chiba-u.jp}
  
${}^4$ 
  E-mail:  {\tt tom@cuphd.nd.chiba-u.ac.jp}
  
${}^5$ 
  E-mail:  {\tt sinohara@graduate.chiba-u.jp}

${}^6$ 
  E-mail:  {\tt shoichi@ei.nagano-nct.ac.jp}

\par 
\par\noindent



\vskip 0.5cm

\newpage
\pagenumbering{roman}




\end{titlepage}


\pagenumbering{arabic}

\baselineskip 14pt
\section{Introduction}

The dual superconductivity in  Yang-Mills vacuum \cite{YM54} is believed to be a promising mechanism \cite{dualsuper}  for quark confinement.  For this picture to be valid, it is supposed that one can extract the relevant Abelian component responsible for duality from non-Abelian gauge theory, i.e.,  Yang-Mills theory,  
since the duality is understood as the electric--magnetic dual in the Abelian gauge theory represented by the Maxwell theory.
Therefore, it is important to give a prescription to extract a variable playing the role of such an Abelian part from the original non-Abelian gauge theory in some sense.
However, such a variable is not necessarily the Abelian gauge field in its appearance.  To emphasize this situation, we use  double quotation marks as ``Abelian''. 

In the previous paper \cite{IKKMSS06}, we have proposed a formulation of Yang-Mills theory on a lattice in a compact form (say, compact lattice formulation).  The compact lattice formulation was constructed so that it reduces in the (naive) continuum limit to the formulation of Yang-Mills theory written in terms of new variables obtained through non-linear change of variables (NLCV) from the original gauge field \cite{KMS05}. The NLCV generates the new variables which have the same form as the Cho--Faddeev--Niemi (CFN) or Cho--Faddeev--Niemi--Shabanov (CFNS) decomposition \cite{Cho80,FN98,Shabanov99}.

Prior to the compact lattice formulation, we have already given another lattice formulation in a non-compact form (referred to as the non-compact lattice formulation hereafter)  \cite{KKMSSI05}. 
The non-compact lattice formulation enabled one to define the magnetic monopole in the {\it gauge invariant} way in Yang-Mills theory on a lattice without introducing fundamental scalar fields.
This is a remarkable result, since the conventional approach of defining the magnetic monopole in Yang-Mills theory without {\it fundamental} scalar fields heavily relies on a specific choice of gauge fixing, the so-called the maximal Abelian gauge (MAG) \cite{KLSW87}, which breaks the color symmetry explicitly in addition to the local gauge symmetry. 
The gauge-invariant magnetic monopole in pure Yang-Mills theory has been constructed by introducing the unit vector field based on NLCV which plays the role of recovering color symmetry broken by a specific choice of color direction according to a Cartan decomposition in MAG. 
However, the magnetic charge resulting from the magnetic monopole defined in this way is not integer-valued in the non-compact formulation.

This drawback was remedied by the subsequent compact formulation \cite{IKKMSS06} which guarantees that the magnetic charge is integer-valued and obeys the Dirac quantisation condition.  
Moreover, the infrared ``Abelian'' dominance and magnetic monopole dominance in the string tension were demonstrated by  numerical simulations in the compact formulation, although such phenomena were found for the first time in the MAG \cite{tHooft81,EI82,SY90,SNW94}.
These results strongly support the dual superconductor picture of QCD vacuum as a promising mechanism of quark confinement.  
However, what is the mechanism for the infrared ``Abelian'' dominance or magnetic monopole dominance   is an unanswered question in this investigation. 
From a theoretical point of view \cite{Kondo06}, on the other hand, it was clarified which variables should be identified with the ``Abelian''part  $\mathbf{V}_\mu$ which is responsible for quark confinement within the continuum formulation \cite{KMS05} so that the variable $\mathbf{V}_\mu$ gives the dominant contribution to the string tension to be calculated from the Wilson loop average (infrared ``Abelian'' dominance). This means that the remaining variable $\mathbf{X}_\mu=\mathbf{A}_\mu-\mathbf{V}_\mu$ decouples in the low-energy or long-distance region to become irrelevant for the string tension, once such an identification of the ``Abelian'' part $\mathbf{V}_\mu$ is achieved.

Our continuum formulation of Yang-Mills theory allows one to introduce the mass term 
$
 \frac12 M_X^2 \mathbf{X}_\mu^2
$
for the remaining field $\mathbf{X}_\mu$ without breaking the local gauge invariance.  
Therefore, the dynamical mass generation for $\mathbf{X}_\mu$ is not prohibited in this formulation. 
If such gluon mass for $\mathbf{X}_\mu$ is generated, it could be a mechanism for infrared ``Abelian'' dominance as pointed out in \cite{Kondo06}. 
The dynamical generation of the gluon mass for $\mathbf{X}_\mu$ yields the decoupling of these degrees of freedom in the low-energy region leaving the ``Abelian'' part $\mathbf{V}_\mu$ as the low-energy modes relevant to quark confinement. 
To confirm this scenario for {\it dynamical Abelian projection} is a main motivation of this paper. 
For MAG, mass generation was so far reported for the off-diagonal gluon component \cite{AS99}.
In this paper, we have measured the mass for the gluon $\mathbf{X}_\mu$ directly by numerical simulations based on our lattice formulation \cite{IKKMSS06}. 
In this paper, moreover, we discuss in detail how to define the lattice variable corresponding especially to the remaining part $\mathbf{X}_\mu$  in the compact lattice formulation to obtain the continuum counterparts to the errors of lattice spacing $\epsilon$. 
Preliminary results have already been reported in \cite{lattice06}.
Finally, it is worth mentioning that the gauge invariance of the Abelian confinement mechanism was also discussed recently in many publications by other group, e.g., \cite{Kanazawa-group}.

\section{Compact formulation}

In order to consider a compact lattice version for the new formulation of Yang-Mills theory, we recall the continuum formulation presented in \cite{KMS05}.
In the continuum formulation \cite{Cho80,KMS05}, we have introduced a color vector field $\vec{n}(x)=(n_A(x))$ $(A=1,2,3)$ of a unit length, i.e., $\vec{n}(x) \cdot \vec{n}(x) := n_A(x)  n_A(x)=1$. 
In what follows, we use the arrow to denote the vector and use the boldface letter to express the Lie-algebra $su(2)$-valued field, e.g., ${\bf n}(x) :=n_A(x)T_A$ with generators of  $su(2)$, $T_A=\frac{1}{2}\sigma_A$ where $\sigma_A$ ($A=1,2,3$)  are Pauli matrices.
The $su(2)$-valued gluon field (gauge potential) $\mathbf{A}_\mu(x)$ is decomposed into two parts:
\begin{align}
  \mathbf{A}_\mu(x) = \mathbf{V}_\mu(x) + \mathbf{X}_\mu(x) ,
  \label{decomp}
\end{align}
in such a way that the color vector field ${\bf n}(x)$ is covariantly constant in the background  field $\mathbf{V}_\mu(x)$:
\begin{align}
 0 = \mathscr{D}_\mu[\mathbf{V}] {\bf n}(x) 
:= \partial_\mu {\bf n}(x) -i g [\mathbf{V}_\mu(x) , {\bf n}(x) ],
 \label{covariant-const}
\end{align}
and that the remaining field $\mathbf{X}_\mu(x)$ is perpendicular to ${\bf n}(x)$:
\begin{align}
  0 = \vec{n}(x) \cdot \vec{X}_\mu(x) \equiv 2{\rm tr}({\bf n}(x)   \mathbf{X}_\mu(x)) .
  \label{nX=0}
\end{align}
Here we have introduced the gauge coupling $g$ and we have adopted the normalization for generators: ${\rm tr}(T_A T_B)= \frac12 \delta_{AB}$. 
Note that $n^A(x)$, $\mathbf{V}_\mu^A(x)$, $\mathbf{X}_\mu^A(x)$ and $\mathbf{A}_\mu(x)$ are real-valued fields and their Lie-algebra forms are Hermitian due to Hermiticity of the generators $T^A$.

By solving the defining equation (\ref{covariant-const}), the   $\mathbf{V}_\mu(x)$ field is obtained in the form:
\begin{align}
  \mathbf{V}_\mu(x) 
  = \mathbf{V}_\mu^{\parallel}(x) + \mathbf{V}_\mu^{\perp}(x)
  = c_\mu(x) {\bf n}(x)   -i g^{-1} [ \partial_\mu {\bf n}(x), {\bf n}(x) ] ,
  \label{Vdef}
\end{align}
where the second term 
$
 \mathbf{V}_\mu^{\perp}(x) := -i g^{-1} [ \partial_\mu {\bf n}(x), {\bf n}(x) ]
 = g^{-1} (\partial_\mu \vec{n}(x) \times \vec{n}(x))_A T_A 
$ 
is perpendicular to ${\bf n}(x)$, i.e., 
$
 \vec{n}(x) \cdot \vec{V}_\mu^{\perp}(x) \equiv 2{\rm tr}({\bf n}(x)   \mathbf{V}_\mu^{\perp}(x)) = 0
$.
Here it should be remarked that the parallel part $\mathbf{V}_\mu^{\parallel}(x)=c_\mu(x) {\bf n}(x)$, $c_\mu(x)= {\rm tr}({\bf n}(x)  \mathbf{A}_\mu(x))$ proportional to ${\bf n}(x)$ can not be determined uniquely from the defining equation (\ref{covariant-const}). 
Imposing the perpendicular condition (\ref{nX=0}) determines $\mathbf{V}_\mu^{\parallel}(x)$ and the remaining part $\mathbf{X}_\mu(x)$  as
\begin{align}
  \mathbf{X}_\mu(x) = -ig^{-1} [ {\bf n}(x), \mathscr{D}_\mu[\mathbf{A}] {\bf n}(x)]     .
  \label{Xdef}
\end{align}
It is easy to check that the sum of $\mathbf{V}_\mu(x)$ and $\mathbf{X}_\mu(x)$ specified respectively by (\ref{Vdef}) and (\ref{Xdef}) agrees with the original field $\mathbf{A}_\mu(x)$ according to (\ref{decomp}).

On a lattice, on the other hand, we introduce the site variable ${\bf n}_{x}$ constructed according to \cite{KKMSSI05}, in addition to the original link variable $U_{x,\mu}$. 
Note that we define a color vector field ${\bf n}_x  :=n^A_x \sigma_A$ on the lattice corresponding to the continuum notation ${\bf n}(x) :=n_A(x)T_A$. 
In this paper, we define the link variable $U_{x,\mu}$ as the exponential of the line integral of a gauge potential $\mathbf{A}_\mu(x)$ along a link from $x$ to $x+\mu$:
\begin{align}
 U_{x,\mu} = \mathscr{P} \exp \left( -ig \int_{x}^{x+\epsilon \mu} dx^\mu \mathbf{A}_\mu(x) \right) ,
 \label{link}
\end{align}
where $\epsilon$ denotes the lattice spacing and $\mathscr{P}$ denotes the path ordering.
In the explicit estimation of the naive continuum limit, 
we adopt in this paper the mid-point definition for the link variable:
\begin{align}
U_{x,\mu} = \exp( -i g \epsilon  \mathbf{A}_\mu(x')) , 
\label{def-U}
\end{align}
using the midpoint $x':=(x+\epsilon \mu/2,\mu)$ of the link $(x,x+\epsilon \mu)$ running from $x$ to $x+\epsilon \mu$.
This prescription is adopted to suppress as much as possible lattice artifacts coming from a finite (nonzero) lattice spacing, in contrast to our previous paper \cite{IKKMSS06} where we have adopted the very naive definition:
$
 U_{x,\mu} = \exp( -i g\epsilon  \mathbf{A}_\mu(x)) . 
$

The link variable $U_{x,\mu}$ and the site variable ${\bf n}_{x}$ transform under the gauge transformation II \cite{KMS05} as
\begin{align}
  U_{x,\mu} \rightarrow \Omega_{x} U_{x,\mu} \Omega_{x+\mu}^\dagger = U_{x,\mu}' , \quad
  {\bf n}_{x} \rightarrow \Omega_{x} {\bf n}_{x} \Omega_{x}^\dagger = {\bf n}_{x}' .
\end{align}
Note that 
${\bf n}_{x}$ is Hermitian, ${\bf n}_{x}^\dagger={\bf n}_{x}$, and $U_{x,\mu}$ is unitary, $U_{x,\mu}^\dagger=U_{x,\mu}^{-1}$.  
It should be remarked that this transformation property follows from the most general form (\ref{link}) for the link variable $U_{x,\mu}$, irrespective of the prescription for the discrete lattice approximation.

The lattice variables $V_{x,\mu}$ and $X_{x,\mu}$ corresponding to $\mathbf{V}_\mu(x)$ and $\mathbf{X}_\mu(x)$ should be expressed in terms of the site variable ${\bf n}_{x}$ and the original link variable $U_{x,\mu}$,
just as the continuum variables $\mathbf{V}_\mu(x)$ and $\mathbf{X}_\mu(x)$ are expressed in terms of ${\bf n}(x)$ and $\mathbf{A}_\mu(x)$, 
However, the definition of lattice variables $V_{x,\mu}$ and $X_{x,\mu}$ is not unique.  They must be defined in a consistent way with the defining equation on a lattice respecting the transformation property.  
We achieve this by solving a lattice version \cite{IKKMSS06} of (\ref{covariant-const}) and (\ref{nX=0}): 
\begin{align}
 {\bf n}_{x} V_{x,\mu}  - V_{x,\mu} {\bf n}_{x+\mu} =0 ,
 \label{Lcc}
\end{align}
\begin{equation}
 {\rm tr}({\bf n}_{x} X_{x,\mu}  ) 
  = 0 .
  \label{cond2m0}
\end{equation} 
The defining equation must be invariant under the gauge transformation, namely, they are form-invariant:
$
 {\bf n}_{x}^\prime V_{x,\mu}^\prime  - V_{x,\mu}^\prime {\bf n}_{x+\mu}^\prime =0 ,
$
and
$
 {\rm tr}({\bf n}_{x}^\prime X_{x,\mu}^\prime  )  = 0 
$.
We identify the lattice variable $V_{x,\mu}$ with a link variable which transforms in the same way as the original link variable $U_{x,\mu}$:
\begin{align}
  V_{x,\mu} \rightarrow \Omega_{x} V_{x,\mu} \Omega_{x+\mu}^\dagger
   = V_{x,\mu}' .
   \label{transf-V}
\end{align}
This requirement guarantees that the defining equation (\ref{Lcc}) is gauge invariant. 
On the other hand, we define the lattice variable $X_{x,\mu}$ so that it transforms in just the same way as the site variable ${\bf n}_x$:
\begin{align}
  X_{x,\mu} \rightarrow \Omega_{x} X_{x,\mu} \Omega_{x}^\dagger
   = X_{x,\mu}' ,
   \label{transf-X}
\end{align}
to realize the adjoint color rotation at the site suggested from the transformation property of the continuum variable. 
By this choice, indeed, the orthogonality condition (\ref{cond2m0}) is kept gauge invariant. 

Explicit construction of the new lattice variables are as follows.
We define $V_{x,\mu}$ as a link variable which is a group element of $G=SU(2)$ related to the  $su(2)$-valued background field $\mathbf{V}_\mu(x)$ through 
\begin{align}
 V_{x,\mu} = \mathscr{P} \exp \left( -ig \int_{x}^{x+\epsilon \mu} dx^\mu \mathbf{V}_\mu(x) \right) .
 \label{link-V}
\end{align}
In the mid-point definition the link variable $V_{x,\mu}$ reads
\begin{align}
  V_{x,\mu} = \exp (-i\epsilon g \mathbb{V}_\mu(x')) ,
\end{align}
where $\mathbb{V}_\mu(x')$ is to be identified with the continuum variable $\mathbf{V}_\mu(x)$ defined by (\ref{Vdef}) in the continuum limit. 
Hence  $V_{x,\mu}$ must be unitary $V_{x,\mu}^\dagger=V_{x,\mu}^{-1}$. 
The same remark as the the link variable $U_{x,\mu}$ for the naive continuum limit holds also for the link variable $V_{x,\mu}$.
In the previous paper \cite{IKKMSS06},  the lattice version (\ref{Lcc}) of the defining equation (\ref{covariant-const}) 
has been solved and the resulting link variable $V_{x,\mu}$ is of the form (up to the normalization) \cite{CGI98}:
\begin{align}
  \tilde{V}_{x,\mu} = \tilde{V}_{x,\mu}[U,{\bf n}] 
  = U_{x,\mu} +  {\bf n}_{x} U_{x,\mu} {\bf n}_{x+\mu} ,
  \label{sol}
\end{align}
and the unitary link variable $V_{x,\mu}[U,{\bf n}]$ has been obtained after  the normalization: 
\begin{eqnarray} 
 V_{x,\mu} = 
V_{x,\mu}[U,{\bf n}] := 
 \tilde{V}_{x,\mu}/\sqrt{{\rm tr} [\tilde{V}_{x,\mu}^{\dagger}\tilde{V}_{x,\mu}]/2} .
\label{cfn-mono-4}
\end{eqnarray}
Indeed, the naive continuum limit $\epsilon \rightarrow 0$ of the link variable (\ref{cfn-mono-4}) reduces to the continuum expression (\ref{Vdef}). 

A naive choice for the lattice variable $X_{x,\mu}$ is given by
$
  U_{x,\mu} V_{x,\mu}^\dagger 
$ 
or
$
V_{x-\mu,\mu}^\dagger U_{x-\mu,\mu} .
$
These are suggested from the relation ${\bf X}_\mu(x)={\bf A}_\mu(x)-{\bf V}_\mu(x)=-{\bf V}_\mu(x)+{\bf A}_\mu(x)$.   In fact, they satisfy the desired transformation property (\ref{transf-X}).  
Note that $V_{x,\mu}^\dagger U_{x,\mu}$ or $
U_{x-\mu,\mu}V_{x-\mu,\mu}^\dagger$ is excluded, since it obeys the adjoint rotation at $x+\mu$ or $x-\mu$, not at $x$.
Then we can construct a lattice variable $X_ {x,\mu}$ as the linear combination:  
\begin{equation}
 \tilde{X}_{x,\mu} = \alpha U_{x,\mu} V_{x,\mu}^\dagger 
 + \beta V_{x-\mu,\mu}^\dagger U_{x-\mu,\mu} ,
  \label{cond2mc}
\end{equation}
to satisfy the desired transformation property (\ref{transf-X}). 
Now we can see that it is reasonable to adopt (\ref{cond2m0}) as a lattice version of the orthogonality equation (\ref{nX=0}).  In fact,  
${\rm tr}({\bf n}_{x} \mathbb{X}_\mu(x))=0$ implies 
(See Appendix for the derivation of (\ref{cond2}))
\begin{align}
 {\rm tr}({\bf n}_{x} \tilde{X}_{x,\mu})
  =&  (\alpha+\beta) {\rm tr}({\bf n}_{x}  \{ {\bf 1}-ig \epsilon  \mathbb{X}_\mu(x) \} ) + (\alpha-\beta){\cal O}(\epsilon^2) 
  \nonumber\\
=& -ig (\alpha+\beta)  \epsilon{\rm tr}({\bf n}_{x} \mathbb{X}_\mu(x) ) + (\alpha-\beta){\cal O}(\epsilon^2) 
= 0 + (\alpha-\beta){\cal O}(\epsilon^2) .
  \label{cond2}
\end{align}
In this way, the lattice variables  $X_{x,\mu}$ is expressed in terms of the site variable ${\bf n}_{x}$ and the original link variable $U_{x,\mu}$ as
\begin{align}
 \tilde{X}_{x,\mu}=\tilde{X}_{x,\mu}[U,\mathbf{n}] 
= (\alpha+\beta) \bm{1} + \alpha U_{x,\mu} {\bf n}_{x+\mu} U_{x,\mu}^\dagger {\bf n}_{x} + \beta {\bf n}_{x} U_{x-\mu,\mu}^\dagger {\bf n}_{x-\mu} U_{x-\mu,\mu}  .
\end{align}
In particular, a good choice is obtained for the symmetric case, i.e., $\alpha=\beta$, since this choice enables us to define the lattice variable $X_{x,\mu}$ so as to reproduce the naive continuum limit of the orthogonality equation up to $O(\epsilon^3)$. 

Finally, we obtain the unitary lattice variable  $\hat{X}_{x,\mu}[U,\mathbf{n}]$ after the normalization:
\begin{align}
 X_{x,\mu}[U,\mathbf{n}] :=& \tilde{X}_{x,\mu}/\sqrt{\mathrm{tr}[\tilde{X}_{x,\mu}^{\dagger}\tilde{X}_{x,\mu}]/2} .
\end{align}
In the numerical simulations, we have adopted the choice $\alpha=\beta$ by the reason mentioned above. 
Moreover, there are some arbitrariness for extracting the Lie-algebra valued variable $\mathbb{X}_\mu$ from the compact lattice variable $\hat{X}_{x,\mu}[U,\mathbf{n}]$. 
This issue will be examined by comparing the results of numerical simulations.


\section{Abelian dominance and gluon mass generation }

\subsection{Identifying the Abelian part and mass term for the remaining part}

The $\mathbf{V}_\mu$  field can be regarded as the ``Abelian'' part  in the reformulated  Yang-Mills theory by the following reasons.
\begin{enumerate}
\item[
i)] The ``Abelian'' part  $\mathbf{V}_\mu^{A}$ corresponds to the diagonal part of the gauge potential $\mathbf{A}_\mu^{A}$ in the context of the conventional MAG which is reproduced when the color vector is aligned in the same direction over the whole spacetime, for example,   
\begin{equation}
 \vec{n}(x) \rightarrow \vec{n}_0 :=(0,0,1) .
\label{MAGlimit}
\end{equation}

\item[
ii)] The Wilson loop average $W(C)$ in Yang-Mills theory written in terms of $\mathbf{A}_\mu^{A}$ is rewritten into the reduced Wilson loop average $\tilde{W}(C)$ which is entirely rewritten in terms of $\mathbf{V}_\mu^{A}$ in the reformulated Yang-Mills theory, as demonstrated in \cite{Kondo06}.

\item[
iii)] The mass term for $\mathbf{X}_\mu^{A}$ can be introduced without breaking gauge invariance in this reformulation \cite{KMS05}.  
In fact, it has been shown to one-loop order \cite{Kondo06} that such an effective mass term is  generated due to the gauge-invariant dimension two condensate $\left< \mathbf{X}_\mu^A \mathbf{X}_\mu^A \right>$ thanks to the {\it gauge invariant} self-interaction term $\frac14 g^2(\epsilon^{ABC}\mathbf{X}_\mu^B \mathbf{X}_\nu^C)^2$ among $\mathbf{X}_\mu$ gluons, in sharp contrast to the ordinary self-interaction term $\frac14 g^2(\epsilon^{ABC}\mathbf{A}_\mu^B \mathbf{A}_\nu^C)^2$ which is not gauge-invariant.  
\end{enumerate}
Therefore, in the energy region lower than the mass $M_X$ of the field $\mathbf{X}_\mu$, the remaining components $\mathbf{X}_\mu$ should decouple or negligible and the $\mathbf{V}_\mu$ field could be dominant.
This leads to the infrared Abelian dominance (in the string tension) in our reformulation. 

Keeping these facts in mind, we proceed to obtain a fitting function of the two--point correlation function.
Suppose that the Yang-Mills theory has the effective mass term:
\begin{equation}
 \frac12 M_X^2 \mathbf{X}_\mu^A \mathbf{X}_\mu^A  .
\end{equation}
An additional quadratic term in $\mathbf{X}_\mu$ of the following type could be generated from  gauge fixing conditions in the differential form \cite{KMS05}.
\begin{equation}
 - \frac{1}{2\beta} (\partial^\mu \mathbf{X}_\mu^A)^2  .
\end{equation} 
This can be understood as follows. 
Recall that we impose an constraint called the new Maximal Abelian gauge (nMAG) to obtain the reformulated Yang-Mills theory with the original gauge symmetry SU(2) even after introducing the color field ${\bf n}(x)$ which apparently increases  gauge degrees of freedom \cite{KMS05}.
Then we introduce a gauge-fixing parameter $\alpha$ for nMAG of the form: 
$- \frac{1}{2\alpha} ((\mathscr{D}^\mu[\mathbf{V}] \mathbf{X}_\mu)^A)^2$.  
This term does not fix the SU(2) gauge invariance. 
Therefore, we adopt the Landau gauge for the overall gauge fixing of $\mathbf{A}_\mu^A$ whose differential form is $\partial_\mu \mathbf{A}_\mu^A=0$.  This gives an additional quadratic term:
$
 - \frac{1}{2\alpha^\prime} (\partial^\mu \mathbf{X}_\mu^A)^2  
$ 
coming from the GF term: 
$
 - \frac{1}{2\alpha^\prime} (\partial^\mu \mathbf{A}_\mu^A)^2  
$.  
Therefore, combining two terms yields an additional term quadratic in $\mathbf{X}_\mu^A$: 
$
 - \frac{1}{2\beta} (\partial^\mu \mathbf{X}_\mu^A)^2  
$
with $\beta^{-1}=\alpha^{-1}+\alpha^{\prime}{}^{-1}$.
Thus we assume the effective propagator for $\mathbb{X}$ gluon of the form:
\begin{align}
 D_{\mu\nu}^{XX}(k) =  \frac{-1}{k^2-M_X^2} \left[ \delta_{\mu\nu} - (1-\beta) \frac{k_\mu k_\nu}{k^2-\beta M_X^2} \right] .
\end{align}
In particular, the limit $\beta \rightarrow \infty$ reproduces the Proca case: 
\begin{equation}
 D_{\mu\nu \infty}^{XX}(k) =  \frac{-1}{k^2-M_X^2} \left[  \delta_{\mu\nu} -\frac{k_\mu k_\nu}{M_X^2} \right] . 
\end{equation}
This form was adopted in the study of off-diagonal gluon mass generation in MAG \cite{AS99} where the mass term $\frac12 M_{\rm off}^2 A_\mu^a A_\mu^a$ was introduced by hand without preserving the gauge invariance.
Note that both nMAG and Landau gauge conditions are exactly satisfied only at $\alpha=0$ and $\alpha^\prime=0$.  
This is realized at $\beta=0$ limit:
\begin{equation}
 D_{\mu\nu 0}^{XX}(k) =  \frac{-1}{k^2-M_X^2} \left[  \delta_{\mu\nu} -  \frac{k_\mu k_\nu}{k^2} \right] ,
 \quad
 D_{\mu\mu 0}^{XX}(k) =  \frac{-(D-1)}{k^2-M_X^2}  
= D_{\mu\mu \infty}^{XX}(k)- \frac{1}{M_X^2} .
\end{equation}
Therefore, the $\beta=0$ limit differs from the previous Proca case used in MAG. 
However, it will turn out below that the constant shift of the propagator gives the same decay  rate and hence the same mass $M_X$ of $\mathbf{X}_\mu$ gluon. 

\subsection{Numerical simulations}

We have generated configurations of link variables $\{U_{x,\mu}\}$ based on the standard heat bath method for the standard
Wilson action. The numerical simulation are performed at $\beta=2.3$, $2.4$ on $24^{4}$ lattice,
at $\beta=2.3$, $2.4$, $2.5$ on $32^{4}$ lattice, 
at $\beta=2.4$, $2.5$, $2.6$ on $36^{4}$ lattice,
and   at $\beta=2.4, 2.5, 2.6$ on $48^4$ lattice by thermalizing 15000 sweeps.  
Here 200 configurations are stored every 300 sweeps. 
Other settings of numerical simulations are the same as those in the previous paper \cite{IKKMSS06}.

\begin{figure}[ptb]
\begin{center}
\vspace{-5mm}%
\includegraphics[width=2.95in]{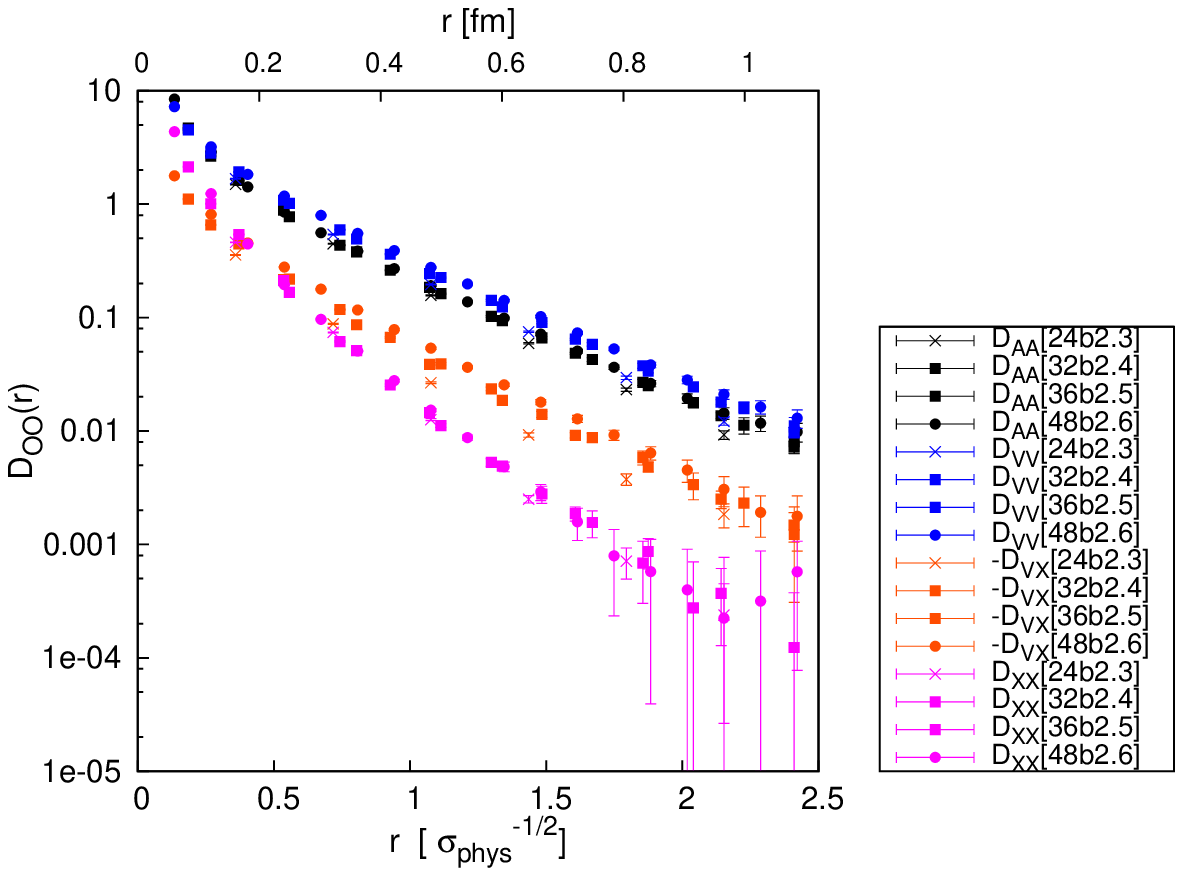}
\includegraphics[width=2.95in]{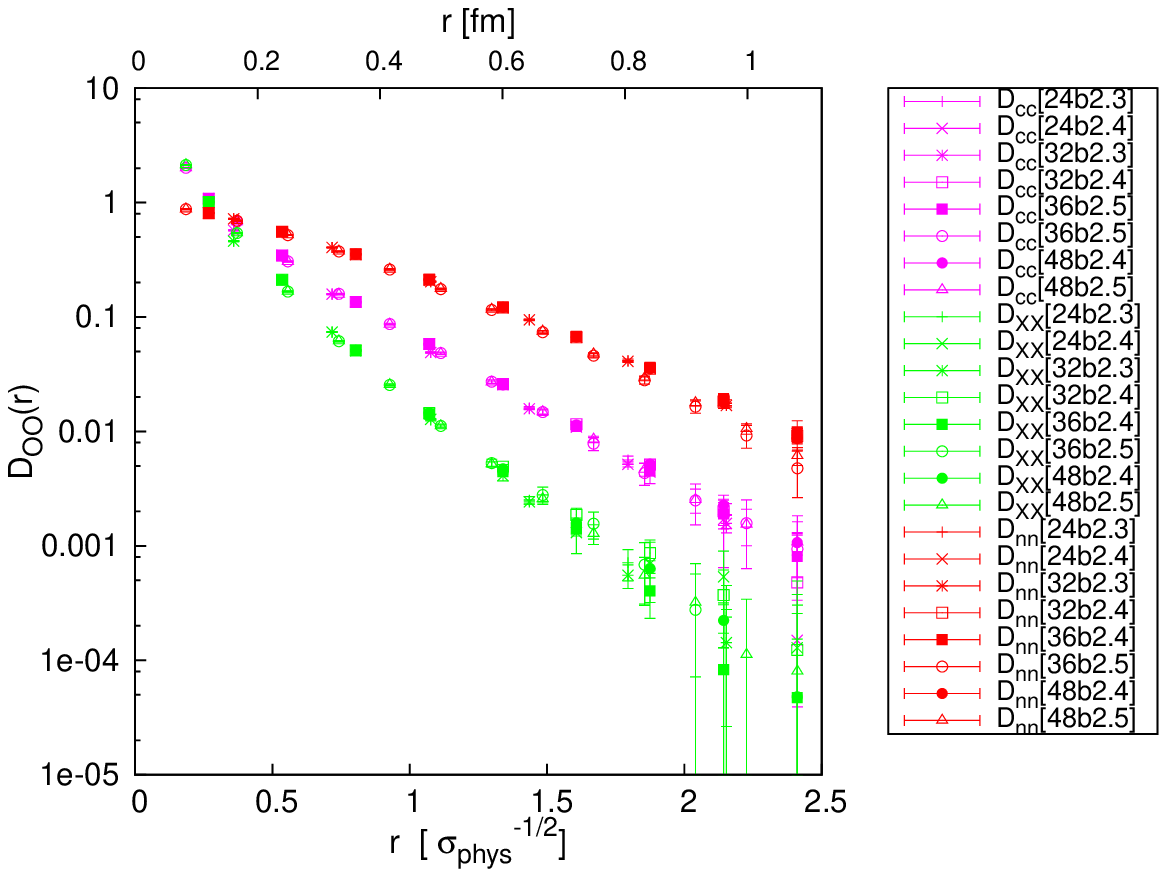}
\vspace{-8mm}
\end{center}
\caption{
Logarithmic plots of scalar-type two-point correlation functions 
$D_{OO^\prime}(r):=\left\langle \mathcal{O}(x) \mathcal{O}^\prime(y) 
\right\rangle$
as a function of the Euclidean distance $r:=\sqrt{(x-y)^2}$  
for  $\mathcal{O}$ and $\mathcal{O}^\prime$. 
(Left panel)
$\mathcal{O}(x)\mathcal{O}^\prime(y)
= \mathbb{V}_\mu^A(x) \mathbb{V}_\mu^A(y)$, 
$\mathbb{A}_\mu^A(x) \mathbb{A}_\mu^A(y)$,  
$-\mathbb{V}_\mu^A(x) \mathbb{X}_\mu^A(y)$,
   $\mathbb{X}_\mu^A(x) \mathbb{X}_\mu^A(y)$, 
(Right panel)
$\mathcal{O}(x)\mathcal{O}^\prime(y)
= {\bf n}^A(x){\bf n}^A(y)$,  
$c_\mu(x)c_\mu(y)$,  $\mathbb{X}_\mu^A(x)\mathbb{X}_\mu^A(y)$,  from above to below
using data on the $24^4$ lattice ($\beta=2.3, 2.4$), 
$32^4$  lattice ($\beta=2.3, 2.4$), $36^4$  lattice ($\beta=2.4, 2.5$), 
and    $48^4$  lattice ($\beta=2.4, 2.5, 2.6$).
Here plots are given in the physical unit [fm] or in unit of square root of the string tension $\sqrt{\sigma_{\rm phys}}$. 
}
\label{fig:prop}%
\end{figure}

We are now ready to study characteristic features of  the reformulated Yang-Mills theory written in terms of new variables $\mathbf{n}^A(x), c_\mu(x), \mathbf{X}_\mu^A(x)$ defined through NLCV of the original field variable $\mathbf{A}_\mu^A(x)$: infrared Abelian dominance, magnetic monopole dominance and  non-vanishing gluon mass. 
Among them, the magnetic monopole dominance in the string tension has already been confirmed in the previous paper \cite{IKKMSS06} using the gauge-invariant magnetic monopole which is guaranteed to have integer-valued magnetic charge subject to the Dirac quantization condition according to our construction of magnetic current based on NLCV.
\footnote{
The proposed NLCV enables one to extract the \textquotedblleft Abelian part\textquotedblright\ $\mathbb{V}_{x,\mu}^A$ irrespective of the choice of the gauge fixing preserving the color
symmetry.  The Yang-Mills theory in the conventional MAG  is reproduced as a very special limit (\ref{MAGlimit}) of our reformulated Yang-Mills theory based on NLCV. 
} 
An advantage of our formulation is that we can confirm such characteristic features for any choice of gauge fixing, not restricted to MAG, since our formulation allows us to take arbitrary type of gauge fixing for the original variable $\mathbf{A}_\mu^A(x)$. 

To study the infrared Abelian dominance and the gluon mass generation in the reformulated Yang-Mills theory, we first define the two-point correlation functions (full propagators) for the independent variables in the new formulation on a lattice, i.e.,  ${\bf n}_{x}^A$, $c_{x,\mu}$ and $\mathbb{X}_{x,\mu}^A$, in addition to the original variable $\mathbb{A}_{x,\mu}^A$. 
For simplicity, we examine just the contracted scalar-type propagator simplified by avoiding the complicated tensor structure: 
\begin{align}
D_{nn}(x-y) =& \left\langle {\bf n}_{x}^A  \ {\bf n}_{y}^A\right\rangle ,
\quad 
D_{cc}(x-y) = \left\langle c_{x',\mu} \  c_{y',\mu}\right\rangle ,
\nonumber\\
D_{XX}(x-y) =& \left\langle \mathbb{X}_{x,\mu}^A \ \mathbb{X}_{y,\mu}^A
\right\rangle , 
\quad
D_{X'X'}(x-y) = \left\langle \mathbb{X}_{x',\mu}^A \ \mathbb{X}_{y',\mu}^A
\right\rangle ,
\end{align} 
and
\begin{equation}
D_{AA}(x-y)=\left\langle \mathbb{A}_{x',\mu}^A \ \mathbb{A}_{y',\mu}^A \right\rangle .
\end{equation} 
Here the Lie-algebra valued gauge potential $\mathbb{A}_{x^{\prime},\mu}$ or $\mathbb{V}_{x^{\prime},\mu}$ is defined from the respective link variable by
\begin{equation}
\mathbb{A}_{x^{\prime},\mu} :=(i/2g\varepsilon) \left[  U_{x,\mu
}-U_{x,\mu}^{\dagger}\right]  , \quad 
\mathbb{V}_{x^{\prime},\mu}= 
(i/2g\varepsilon) \left[  V_{x,\mu}-V_{x,\mu}^{\dagger}\right]  .
\end{equation} 
For the variable $\mathbb{X}_{x,\mu}$, on the other hand, we examined two options:
 one is extracted from decomposing the gauge potential (group-valued): 
\begin{equation}
\mathbb{X}_{x,\mu} :=(i/2g\varepsilon) \left[  X_{x,\mu}-X_{x,\mu}^{\dagger}\right]  ,
\label{Xpro-1}
\end{equation} 
and the other is from the definition of the decomposition (Lie-algebra-valued):
\begin{equation}
\mathbb{X}_{x^{\prime},\mu} :=\mathbb{A}_{x^{\prime},\mu}-\mathbb{V}%
_{x^{\prime},\mu}.
\label{Xpro-2}
\end{equation} 
The field $c_{x',\mu}$ is defined by 
\begin{equation}
 c_{x',\mu} :={\rm tr}( \mathbf{n}_{x} V_{x,\mu})={\rm tr}( V_{x,\mu}\mathbf{n}_{x+\mu}) .
\end{equation}

\begin{figure}[ptb]
\begin{center}
\vspace{-5mm}%
\includegraphics[width=3.5in]{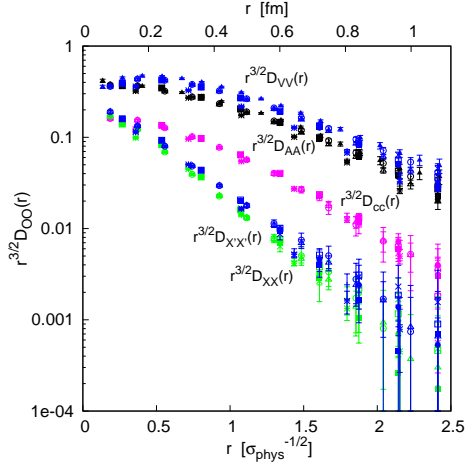}
\vspace{-8mm}
\end{center}
\caption{
Logarithmic plots of the rescaled correlation function 
$r^{3/2}D_{OO}(r)$ as a function of $r$ for $O=\mathbb{V}_\mu^A,\mathbb{A}_\mu^A,c_\mu,\mathbb{X}_\mu^A$ (and $\mathbb{X}^\prime{}_\mu^A$)  from above to below, using the same colors and symbols as those in Fig.~\ref{fig:prop}.
Here two sets of data for the correlation function $D_{XX}(x-y)$ are plotted according to the two definitions (\ref{Xpro-1}) and (\ref{Xpro-2}) of the $\mathbb{X}_\mu^A$ field on a lattice.  
}
\label{fig:rescaled-propagator}%
\end{figure}

The numerical results are presented in Fig.~\ref{fig:prop}.
As is quickly observed from the left panel of Fig.~\ref{fig:prop}, 
$D_{VV}(x-y)$ and $D_{AA}(x-y)$ exhibit quite similar behaviors in the measured range of the Euclidean distance $r=|x-y|:=\sqrt{(x-y)^2}$.
In order to determine the physical scale, we have used the relationship between the (inverse) gauge coupling $\beta$ and lattice spacing $\epsilon$ given in \cite{Kato} 
\footnote{
We use the relationship between the physical units, 
$1 {\rm GeV}^{-1}=0.197327 {\rm fm}$ or $1 {\rm GeV}=5.06773 {\rm fm}^{-1}$.  
This comes from $\hbar c=0.197327 {\rm GeV} \cdot {\rm fm} $.
}
which is summarized in Table~\ref{table:dictionary}.

\begin{table}
\begin{center}%

\begin{tabular}
[c]{||l|c|c|c|c|c|c||}\hline\hline
& \multicolumn{2}{|c|}{lattice spacing $\epsilon$} &
\multicolumn{4}{||c||}{lattice size $L$ $\mathrm{[fm]}$}\\\hline
\multicolumn{1}{||c|}{$\beta$} & { $[1/\sqrt{\sigma_{phys}}]$ } & {
$\mathrm{[fm]}$ } & \multicolumn{1}{||c|}{$24^{4}$} & $32^{4}$ & $36^{4}$ &
$48^{4}$\\\hline\hline
\multicolumn{1}{||c|}{$2.3$} & $0.35887$ & $0.1609$ & $3.8626$ & $5.1501$ &
$5.7939$ & $7.7252$\\\hline
\multicolumn{1}{||c|}{$2.4$} & $0.26784$ & $0.1201$ & $2.8828$ & $3.8438$ &
$4.3242$ & $5.7657$\\\hline
\multicolumn{1}{||c|}{$2.5$} & $0.18551$ & $0.08320$ & $1.9967$ & $2.6622$ &
$2.9950$ & $3.9934$\\\hline
\multicolumn{1}{||c|}{$2.6$} & $0.13455$ & $0.06034$ & $1.4482$ & $1.9309$ &
$2.1723$ & $2.8964$\\\hline\hline
\end{tabular}

\end{center}
\caption{
The lattice spacing $\epsilon$ and the lattice size $L$ of the lattice volume $L^4$  at various value of  $\beta$ in the physical unit [fm] and the unit given by $\sqrt{\sigma_{phys}}$. 
}
\label{table:dictionary}
\end{table}

From the right panel of Fig.~\ref{fig:prop}, 
$D_{VV}(x-y)$  ($D_{nn}(x-y)$ or $D_{cc}(x-y)$) is dominant compared to $D_{XX}(x-y)$ which decreases more rapidly than other correlation functions in $r$.  
This implies the infrared ``Abelian'' dominance, provided that the components $\mathbf{V}_\mu^A(x)$ composed of ${\bf n}^A_{x}$ and $c_{x,\mu}$ are identified with the ``Abelian'' part of $\mathbf{A}_\mu^A(x)$.  
As is seen from the left  panel of Fig.~\ref{fig:prop},  a non-trivial mixed correlation function
$\left\langle \mathbb{V}_\mu^A(x') \mathbb{X}_\mu^A(y') \right\rangle < 0$
exists, since $\mathbb{V}_\mu^A(x)$ includes a perpendicular component to ${\bf n}^A(x)$. 

Fig.~\ref{fig:prop} demonstrates nice independence of our results against variations of the  ultraviolet cutoff (the lattice spacing $\epsilon$).  The propagators calculated at the lattices with different $\epsilon$ follow the same curve if plotted in the physical units. These accurate plots provide an additional support that the results presented in this paper are definitely not lattice artifacts.

Note that we must impose the gauge fixing condition for the original variable $\mathbb{A}_{x^{\prime},\mu}$  to obtain the correlation function.  In our simulations,  we have chosen the lattice Landau gauge (LLG) for the original field $\mathbf{A}_\mu^A(x)$ for this purpose.  
Thus we have confirmed the infrared ``Abelian'' dominance with color symmetry being kept, since the Landau gauge keeps the color symmetry. 
This is one of our main results. 
The infrared Abelian dominance was so far obtained only for the MAG which breaks the color symmetry explicitly.   
As already mentioned, moreover, we can choose any other gauge and  we can study using this formulation if the infrared ``Abelian'' dominance can be observed in any other gauge. 
We hope we can report the results in the other gauge in  future investigations.

Next, we determine the gluon mass generated in the non-perturbative way by examining the correlation functions in more detail. 
The gauge boson propagator $D_{\mu\nu}^{XX}(x-y)$ is related to the Fourier transform of the massive propagator $D_{\mu\nu}^{XX}(k)$: 
\begin{equation}
 D_{\mu\nu}^{XX}(r)
=\left\langle \mathbb{X}^A_{\mu}(x)  \mathbb{X}^A_{\nu}%
(y)\right\rangle =\int\frac{d^{4}k}{(2\pi)^{4}}e^{ik(x-y)} 
D_{\mu\nu}^{XX}(k).
\end{equation}
Then the scalar-type propagator $D^{XX}(r):=D^{XX}_{\mu\mu}(x)$  as a function of $r$  should behave for large  $M_X r$ as  
(See \cite{AS99} for details of the integral calculation.)
\begin{equation}
D_{XX}(r)
=\left\langle \mathbb{X}^A_{\mu}(x)  \mathbb{X}^A_{\mu}(y)\right\rangle 
=\int\frac{d^{4}k}{(2\pi)^{4}}e^{ik(x-y)} \frac{3}{k^{2} +M_X^{2}}
\simeq\frac{3\sqrt{M_X}} {2(2\pi)^{3/2}} \frac{e^{-M_X r}}{r^{3/2}} .
\end{equation}
Therefore, the scaled propagator $r^{3/2}D_{XX}(r)$ should be  proportional to $\exp(-M_X r)$ for  $M_X r \gg 1$ with $M$ being the damping rate of $r^{3/2}D_{XX}(r)$.  In other words, the mass $M_X$ of the gauge field $\mathbb{X}_{\mu}$ can be estimated from the slope in the logarithmic plot of the scaled propagator $r^{3/2}D_{XX}(r)$ as a function of $r$.
\footnote{
Here we have assumed that the anomalous dimension is sufficiently small so that the exponent of the power of $r$ is the same as the tree value. 
}


\begin{figure}[ptb]
\begin{center}
\vspace{-16mm}%
\includegraphics[width=2.95in]{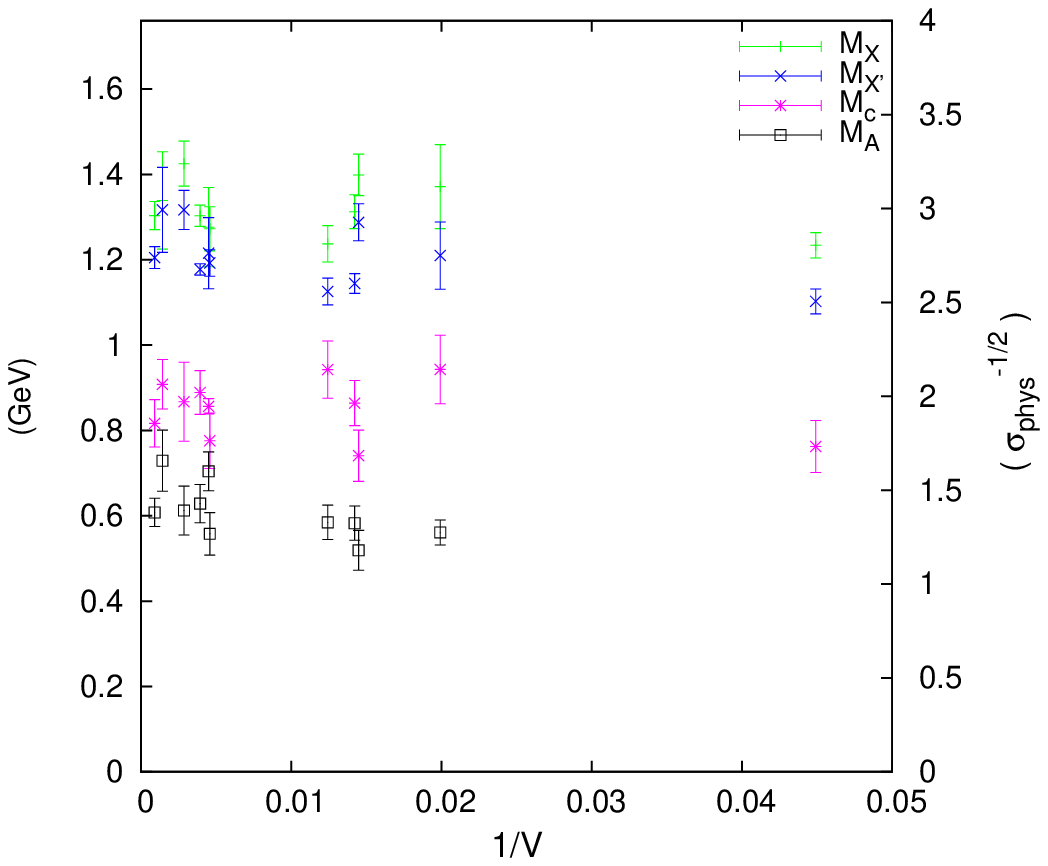}
\includegraphics[width=2.95in]{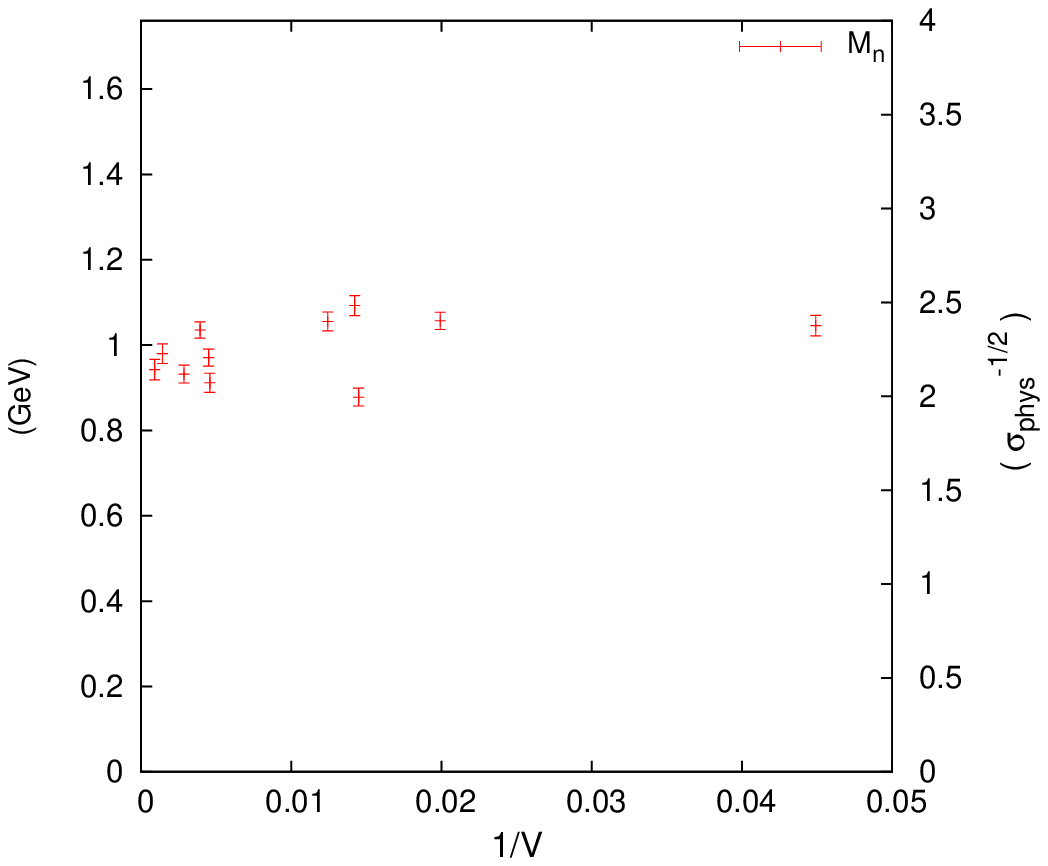}
\vspace
{-8mm}
\end{center}
\caption{
Gluon ``mass'' and decay rates (in units of GeV and $\sqrt{\sigma_{\rm phys}}$) as the function of the inverse lattice volume $1/V$ in the physical unit. 
(Left panel) for $\mathcal{O}=\mathbb{X}_\mu^A,  (\mathbb{X}^\prime{}_\mu^A), c_\mu, \mathbb{A}_\mu^A$ from above to below extracted according to the fitting: $\left\langle \mathcal{O}(x) \mathcal{O}(y) \right\rangle \sim r^{-3/2} \exp (-M_{\mathcal{O}}r)$, 
(Right panel) for $\mathbf{n}^A(x)$ extracted according to the fitting: $\left< \mathbf{n}^A(x) \mathbf{n}^A(y) \right> \sim \exp (-M_n r)$. 
}
\label{fig:mass}%
\end{figure}

Fig.~\ref{fig:rescaled-propagator}  shows the logarithmic plots of the scaled scalar-type propagator for 
$\mathbb{A}_{x^{\prime},\mu}$, 
$c_{x',\mu}$
and 
$\mathbb{X}_{x,\mu}$
as a function of the distance $r$  measured in the physical unit [fm] and in  unit of square root of the string tension $\sqrt{\sigma_{\rm phys}}=440$ MeV. 
According to Fig.~\ref{fig:rescaled-propagator},  we find just small difference between two types of $D_{XX}(x-y)$ defined by (\ref{Xpro-1}) or (\ref{Xpro-2}) over several choices of lattice spacing  (i.e., several values of $\beta$, $\beta=2.3, 2.4, 2.5, 2.6$).
Therefore, we can use either definition of the lattice variable 
$\mathbb{X}_{x^{\prime},\mu}$ to obtain $D_{X'X'}(x-y)$ in the consistent manner.

In Fig.~\ref{fig:mass}, the measured values for the gluon mass are plotted as the function of the inverse lattice volume $1/V$ in the physical unit, to study the finite-size effect on the mass. 
The finite lattice-size effect seems to be small for the gluon mass $M_X$. Here the error bars originate from the fitting procedure for obtaining the slope, but no systematic errors such as finite-volume are included. 
In this way, we have estimated the mass for the $\mathbb{X}$ gluon:
\begin{align}
  M_{X} &\simeq 2.98  \sqrt{\sigma_{\rm phys}} \simeq   
1.31 
  {\rm GeV} ,
  \nonumber\\
  M_{X'} &\simeq 2.69  \sqrt{\sigma_{\rm phys}} \simeq  
1.19 
 {\rm GeV} .
\end{align}

Even after the whole gauge fixing, our formulation preserves color symmetry in sharp contrast to the conventional MA gauge. 
In view of the fact that our reformulated Yang-Mills theory reproduces the Yang-Mills theory in MA gauge as a special limit, the remaining part $\mathbb{X}_\mu^A(x)$ could correspond to the  off-diagonal part in this limit. 
From this point of view, our result is consistent with the result obtained for the off-diagonal gluon mass in MAG \cite{AS99}.

Moreover, we have simultaneously estimated the decay rate for the new fields 
${\bf n}^A(x)$, $c_{\mu}(x)$, $\mathbb{V}_{\mu}^A(x)$
and the original gauge field 
$\mathbb{A}_{\mu}^A(x)$
by imposing the LLG as the overall gauge fixing. 
For $\mathcal{O}=\mathbb{X}_\mu^A, c_\mu, \mathbb{A}_\mu^A, \mathbb{V}_{\mu}^A$,  the decay rate $M_{\mathcal{O}}$ is extracted according to the fitting: $\left\langle \mathcal{O}(x) \mathcal{O}(y) \right\rangle \sim r^{-3/2} \exp (-M_{\mathcal{O}}r)$. 
Fig.~\ref{fig:mass} indicates not so small finite volume effect for data of $1/V > 0.02$.   
Using the data of $1/V<0.02$, therefore, we have estimated the decay rate (or ``mass'') as
\begin{align}
  M_{n} &\simeq 2.24  \sqrt{\sigma_{\rm phys}} 
\simeq 0.986   {\rm GeV} ,
  \nonumber\\
  M_{c} &\simeq 1.94  \sqrt{\sigma_{\rm phys}}
\simeq 0.856  {\rm GeV} ,
  \nonumber\\
  M_{A} &\simeq 1.35  \sqrt{\sigma_{\rm phys}}
\simeq 0.596  {\rm GeV} .
\end{align}
The decay rate $M_c$ obtained from the correlation function of 
$c_\mu(x)$
 field is slightly larger than that expected from the result in MAG. 
It should be remarked that the decay rate for the correlation function of $\mathbf{n}^A(x)$ field is extracted according to the fitting function 
$\left< \mathbf{n}^A(x) \mathbf{n}^A(y) \right> \sim \exp (-M_n r)$ which is not yet justified from the theoretical consideration.  This might be an origin of the large value of $M_n$. 
More simulations on the larger lattice are expected to  eliminate finite volume effect for these values. 
However, we have no argument for guaranteeing the gauge invariance of these values or for identifying these values with their ``masses''.  
In fact, the field $c_\mu(x)$ is not gauge invariant.  
These issues will be checked in further investigations based on our reformulation. 


Finally, we comment on the \textquotedblleft Abelian\textquotedblright part 
$\mathbb{V}_{\mu}^A(x)$,
since our treatment of the \textquotedblleft Abelian\textquotedblright part 
$\mathbb{V}_{\mu}^A(x)$
is different from the conventional approach based on MAG. 
The above result yields the ``mass'' of the \textquotedblleft Abelian\textquotedblright part 
$\mathbb{V}_{\mu}^A(x)$
:
$M_{V} \simeq M_{A} \simeq  0.59 {\rm GeV}$
. 
This value is nearly equal to that of the diagonal gluon mass obtained by imposing the Landau gauge in the conventional approach as reported in the second paper of \cite{AS99} where the Landau gauge was imposed on the Abelian diagonal part $a_\mu(x)$ in addition to the MAG for off-diagonal gluon field $A^a_\mu(x)$ defined by the Cartan decomposition
$\mathbf{A}_\mu(x)=A_\mu^a(x) T^a + a_\mu(x) T^3$ ($a=1,2$).
Therefore, the prescription of gauge fixing in \cite{AS99} is different from ours.


\section{Conclusion and discussion}

In this paper we have developed a compact lattice formulation of  SU(2) Yang-Mills theory proposed in the previous paper \cite{IKKMSS06} as the lattice version of the NLCV which was once called the CFN or CFNS decomposition. This resolves all drawbacks of the previous non-compact lattice formulation of our own \cite{KKMSSI05}.%
\footnote{
This is done up to specifying the integration measure for the respective new variable.  It was not necessary to resolve this issue for obtaining the results reported in this paper.
}
This compact formulation has enabled one to define the gauge-invariant magnetic monopole with the magnetic charge subject to Dirac quantisation condition and to extract the \textquotedblleft Abelian\textquotedblright\ part $\mathbb{V}_\mu(x)$ yielding the infrared ``Abelian'' dominance in the string tension for any choice of the gauge fixing for the original gauge field  $\mathbb{A}_\mu(x)$ in the original YM theory. 
 
In order to confirm the dynamical mass generation for the remaining part $\mathbb{X}_\mu(x)$ as a mechanism for the infrared ``Abelian'' dominance, we have measured the two-point correlation function  (the full propagator in real space) in our lattice formulation by imposing LLG for the original gauge field  $\mathbb{A}_\mu(x)$ as the whole gauge fixing. 
We have found the infrared ``Abelian'' dominance in the sense that the
$\mathbb{X}_\mu(x)$
propagator is suppressed in the long distance compared to 
${\bf n}(x)$ and $c_\mu(x)$
 (and 
$\mathbb{V}_\mu(x)$
propagators as an immediate consequence of dynamically generated mass $M_X=1.2 \sim 1.3 ~{\rm GeV}$ for 
$\mathbb{X}_\mu(x)$
(which is larger than the decay rate of other gluon field propagators).

Even after the whole gauge fixing, our formulation can preserve color symmetry by choosing the gauge-fixing condition which does not break color symmetry, e.g., Landau gauge.  This opens a path to examine  color confinement in the same framework as quark confinement in the dual superconductivity picture.  This feature is in sharp contrast to the conventional MA gauge breaking color symmetry, although our formulation reproduces the MA gauge as a special limit (\ref{MAGlimit}). 
It is important to demonstrate explicitly the gauge-fixing independence of our results obtained in this paper for  establishing the gauge-invariant mechanism for quark confinement. 

\section*{Acknowledgments}
The numerical simulations have been done on a supercomputer 
(NEC SX-5 and NEC SX-8R)
at Research Center for Nuclear Physics (RCNP), Osaka University. This project is
also supported 
by the Large Scale Simulation Program No.06-17 (FY2006) and No.07-15 (FY2007)
of High Energy Accelerator Research Organization (KEK). 
This work is financially supported by 
Grant-in-Aid for Scientific Research (C)18540251 from 
JSPS
and in part by Grant-in-Aid for Scientific Research on Priority Areas (B)13135203 from MEXT.

\appendix
\section{The accuracy of the naive continuum limit}

For the naive continuum limit $\epsilon \rightarrow 0$, we show that the  lattice variable defined by 
$
X_{x,\mu} := \alpha U_{x,\mu} V_{x,\mu}^\dagger 
 +  \beta V_{x-\mu,\mu}^\dagger U_{x-\mu,\mu}
$ 
yields 
$\mathbb{X}_{x,\mu}=\mathbf{X}_{\mu}(x)+{\cal O}(\epsilon^{2})$
for $\alpha=\beta$, 
while 
$\mathbb{X}_{x,\mu}=\mathbf{X}_{\mu}(x)+{\cal O}(\epsilon)$
for $\alpha\not=\beta$.  
The repeated use of the Baker-Campbell-Hausdorff formula yields
\begin{align}
& U_{x,\mu}\hat{V}_{x,\mu}^{\dagger} \ \text{or} \  
\hat{V}_{x-\mu,\mu}^{\dagger}U_{x-\mu,\mu}
\nonumber\\
& = \exp (-ig\epsilon\mathbb{A}_{x + \mu/2,\mu}) \exp (ig\epsilon\mathbb{V}_{x + \mu/2,\mu}) \ \text{or} \ 
\exp (ig\epsilon\mathbb{V}_{x - \mu/2,\mu}) \exp (-ig\epsilon\mathbb{A}_{x - \mu/2,\mu}) 
\nonumber\\
&  =\exp\left\{  -ig\epsilon\mathbb{X}_{x \pm \mu/2,\mu}  \pm  \frac{\left(  g\epsilon\right)^{2}}{2}\left[  \mathbb{A}_{x \pm \mu/2,\mu},\mathbb{V}_{x \pm \mu/2,\mu}\right]  +{\cal O}(\epsilon^{3})\right\}  
\nonumber\\
&  =\exp\left\{  -ig\epsilon\mathbb{X}_{x,\mu} \mp i\frac{g\epsilon^{2}}{2}\partial_{\mu}\mathbb{X}_{x,\mu}  \pm \frac{\left(  g\epsilon\right)^{2}}{2}\left[  \mathbb{A}_{x \pm \mu/2,\mu},\mathbb{V}_{x \pm \mu/2,\mu}\right]
+{\cal O}(\epsilon^{3})\right\}  
\nonumber\\
&  =\exp(-ig\epsilon\mathbb{X}_{x,\mu})\exp\left\{ \mp i\frac{g\epsilon^{2}}{2}\partial_{\mu}\mathbb{X}_{x,\mu} \pm \frac{\left(  g\epsilon\right)^{2}}{2}\left[  \mathbb{A}_{x,\mu},\mathbb{V}_{x,\mu}\right]  +{\cal O}(\epsilon
^{3})\right\}  
\nonumber\\
&  =\exp(-ig\epsilon\mathbb{X}_{x,\mu})\left[  \bm{1} \mp i\frac{g\epsilon^{2}}{2}\partial_{\mu}\mathbb{X}_{x,\mu} \pm \frac{\left(  g\epsilon\right)  ^{2}}{2}\left[  \mathbb{A}_{x,\mu}, \mathbb{V}_{x,\mu}\right]  +{\cal O}(\epsilon^{3})\right]  ,
\end{align}
where we have used 
$
\mathbb{X}_{x,\mu}:=\mathbb{A}_{x,\mu
}-\mathbb{V}_{x,\mu}
$, 
$
\mathbb{V}_{x\pm\mu/2,\mu}=\mathbb{V}_{x,\mu} 
\pm\epsilon/2\partial_{\mu}\mathbb{V}_{x,\mu}
$,   
$
\mathbb{X}_{x\pm\mu/2,\mu
}=\mathbb{X}_{x,\mu}\pm\epsilon/2\partial_{\mu}\mathbb{X}_{x,\mu}
$,
and
$\mathbb{A}_{x\pm\mu/2,\mu}=\mathbb{A}_{x,\mu}%
\pm\epsilon/2\partial_{\mu}\mathbb{A}_{x,\mu}$.
Thus we have 
\begin{align}
&  \alpha U_{x,\mu}\hat{V}_{x,\mu}^{\dagger}+\beta\hat{V}_{x-\mu,\mu}%
^{\dagger}U_{x-\mu,\mu}
\nonumber\\
&  =\exp(-ig\epsilon\mathbb{X}_{x,\mu})\left[  (\alpha+\beta) \bm{1} +(\alpha-\beta)\left\{  -i \frac{g\epsilon^{2}}{2}\partial_{\mu}\mathbb{X}_{x,\mu}+\frac{\left(  g\epsilon\right)  ^{2}}{2}\left[  \mathbb{A}_{x,\mu}, \mathbb{V}_{x,\mu}\right]  \right\}  +{\cal O}(\epsilon^{3})  \right]  
\nonumber\\
&  =(\alpha+\beta) \exp \left[  -ig\epsilon\mathbb{X}_{x,\mu} +(\alpha-\beta)/(\alpha+\beta) \left\{  -i \frac{g\epsilon^{2}}{2}\partial_{\mu}\mathbb{X}_{x,\mu}+\frac{\left(  g\epsilon\right)  ^{2}}{2}\left[  \mathbb{A}_{x,\mu}, \mathbb{V}_{x,\mu}\right]  \right\}  +{\cal O}(\epsilon^{3})  \right] .
\end{align}
The statement follows from the fact that the choice $\alpha=\beta$ eliminates  order $\epsilon^2$ terms. 



\end{document}